\documentclass[twocolumn, 10pt]{article}
\usepackage[utf8]{inputenc}
\usepackage{mathptmx}
\usepackage[affil-it]{authblk} 
\usepackage[a4paper, margin=1in]{geometry}
\usepackage{sectsty}
\usepackage{titlesec}
\usepackage{indentfirst}
\usepackage{array,multirow}
\usepackage{fancyhdr}
\usepackage{graphicx}
\usepackage{caption}
\usepackage{float}

\title{\LARGE \bf
What Timing for an Automated Vehicle\\to Make Pedestrians Understand Its Driving Intentions\\for Improving Their Perception of Safety?
}

\author{
        \large \textbf{Hailong Liu~$^{*~1}$, Takatsugu Hirayama~$^{\#}$, Luis Yoichi Morales~$^{\#}$, Hiroshi Murase~$^{*}$}\\  
        $^{*}$~Graduate School of Informatics, Nagoya University\\
        $^{\#}$~Institutes of Innovation for Future Society, Nagoya University\\
        \normalsize
        Furo-cho, Chikusa-ku, Nagoya, Aichi, 464-8601, JAPAN\\
        $^{1}$~E-mail: lhl881210@live.com
        }
\date{}

\begin{document}
\maketitle
\thispagestyle{empty}
\pagestyle{empty}

\begin{abstract}
Although automated driving systems have been used frequently, they are still unpopular in society.
To increase the popularity of automated vehicles (AVs), assisting pedestrians to accurately understand the driving intentions and improving their perception of safety when interacting with AVs are considered effective.
Therefore, the AV should send information about its driving intention to pedestrians when they interact with each other.
However, the following questions should be answered regarding how the AV sends the information to them:
1) What timing for an AV to make pedestrians understand its driving intentions after being noticed by them?
2) What timing for an AV to make pedestrians feel safe after being noticed by them?
Thirteen participants were invited to interact with a manually driven vehicle and an AV in an experiment.
The participants' gaze information and a subjective evaluation of their understanding of the driving intention as well as their perception of safety were collected.
By analyzing the participants' gaze duration on the vehicle with their subjective evaluations, we found that the AV should enable the pedestrian to accurately understand its driving intention within $0.5 \sim 6.5$~[s] and make the pedestrian feel safe within $0.5\sim8.0$~[s] while the pedestrian is gazing at it.
\end{abstract}

\section{INTRODUCTION}
As the development of automated driving technology has progressed, it has been used in a variety of cases such as intelligent traffic systems, goods distribution~\cite{jennings2019study}, and hospital logistics~\cite{niechwiadowicz2008robot}.
As a result, interaction between people and automated vehicles (AVs) is expected to increase.
This new technology is often not accepted by the public at the current stage of popularization within society~~\cite{Hancock2019} due to a lack of trust in AVs~\cite{Pettigrew2019}.
The essential reason for this distrust is the fear of the unknown~\cite{mcallister2017concrete}, specifically a lack of knowledge regarding the intended actions of the AV while driving, i.e., the current and subsequent actions that would be taken by the AV.
Therefore, many studies claim that providing information to pedestrians regarding the driving intention of the AV is helpful in improving pedestrians' understanding of driving intentions and their perception of safety in interactions~\cite{Habibovic2018,rasouli2019}.
Communication between pedestrians and the AV can improved the understanding of driving intentions and perception of safety are considered effective in increasing the popularity of AVs in society.
 
Thus, in this work the following two problems for pedestrian--vehicle interaction are studied:
\begin{enumerate}
\item What timing for an AV to make pedestrians understand its driving intentions after being noticed by them?
\item What timing for an AV to make pedestrians feel safe after being noticed by them?
\end{enumerate}

For the above questions, we formulate and propose a hypothesis based on a decision--making process of pedestrians, including the situation model and the theory of risk homeostasis.
Based on this hypothetical model, we design an experiment of pedestrian--vehicle interaction.
The participants' gaze information, and their subjective evaluations of the understanding of driving intentions and their perception of safety, are collected.
We seek to identify when pedestrians do not understand the intention of the vehicle, as well as when pedestrians feel danger, by analyzing the participants' gaze duration on the vehicle with their subjective evaluations.

\section{RELATED WORKS}
\begin{figure*}[tb]
\centering
\includegraphics[width=0.89\linewidth]{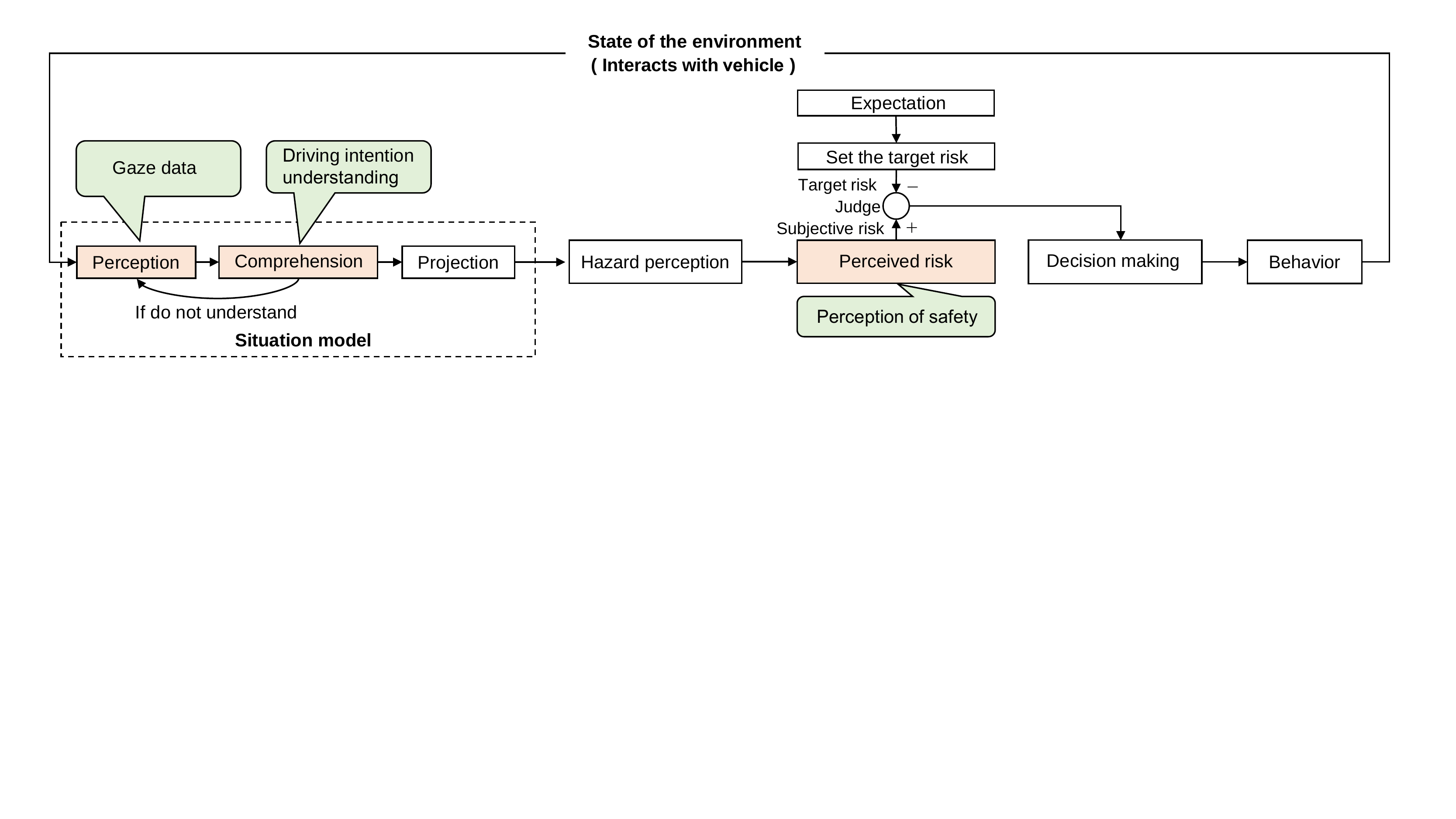}
\caption{Decision--making process of a pedestrian including the situation model and the theory of risk homeostasis.}
\label{fig:model}
\vspace{-5mm}
\end{figure*}

In most studies of pedestrian--vehicle interaction, the availability of different information transmission methods has been subjectively evaluated.
Stefanie et al. evaluated the communication efficacy of external human--machine interfaces (eHMIs) by various light signals with the use of questionnaires and interviews~\cite{Stefanie2019}.
Clercq et al. asked participants to continuously evaluate their feeling of safety by pressing a button during a pedestrian--AV interaction~\cite{Clercq2019}.

In addition, the pedestrians' gaze behavior could be used as an objective factor to analyze pedestrian and vehicle interactions.
This is explained by the fact that the observation of the vehicle by the pedestrian could be considered as his/her desire to obtain information from the vehicle, e.g., determining the intent of the vehicle and predicting if the interaction is dangerous.
For example, Dey et al. found that the gaze point of pedestrians gradually gathered in the driver's position through a windshield when a manually driven vehicle (MV) was approaching~\cite{dey2019gaze}.
In our previous study~\cite{liu2020gaze}, we found that there was a correlation between pedestrians' gaze durations and their understanding of the driving intentions of the AV.
Besides, we considered that pedestrians' gaze durations on the AV could represent the request for information about the AV.
Therefore, we suggested that the AV should send information about its driving intentions to the pedestrian when they interact with each other.

However, when AVs should send information to pedestrians is still an unsolved issue.
To solve this issue, we analyze changes in pedestrians' understanding of driving intentions and changes in their perception of safety according to the gaze duration when they are interacting a AV in this paper.

\section{DECISION--MAKING MODEL OF PEDESTRIAN}
To improve pedestrian's perception of safety during interactions with AVs, the mechanism upon which the perception of safety is based should be clarified.
Thus, we propose a hypothesis which shows the decision--making process of a pedestrian who is interacting with a vehicle.
It is showed in Fig.~\ref{fig:model}.
This hypothesis includes three parts: situation awareness, hazard perception, and decision-making based on risk homeostasis.

Situation awareness could be modeled by the situation model~\cite{endsley2017toward}.
Firstly, situation model relies on the perception of things in the surrounding environment, e.g., AV's relative position, relative distance, and relative speed.
Secondly, comprehension shows the understanding of the current state of the AV in a given situation, such as the driving intention of the AV.
Thirdly, based on the result of comprehension, the pedestrian will predict the driving behavior and moving trajectory of the AV.

After establishing situation awareness, the pedestrian realizes hazards, such as anomaly detection by comparing the predicted driving behavior of the AV with his/her experience.
The subjective risk is assessed by evaluating the perceived hazards.
Subsequently, the subjective risk could be seen as the degree to which the pedestrian feels threatened, e.g., the perception of safety or the perception of danger.

The pedestrian decides his/her behavior by comparing the perceived risk with his/her acceptable risk level according to the risk homeostasis theory~\cite{wilde1982theory}.
If the perceived risk is lower than the acceptable risk level,
the pedestrian will make a decision to do a riskier behavior.
In the opposite case, the pedestrian will make a decision to change the behavior become more careful.

According to the above hypothesis, the gaze duration of the pedestrian at the AV will increase during interaction if pedestrians do not clearly understand the driving intentions of the AV~\cite{liu2020gaze}.
Besides, the pedestrians may also perceive further danger due to a lack of the understanding of driving intentions.
Thus, the gaze duration could be used to objectively evaluate the pedestrians' understanding of the driving intentions and their perception of safety in an interaction with the AV.

\section{EXPERIMENT DESIGN}

\begin{figure}[t]
\centering
\begin{minipage}[t]{0.495\linewidth}
\centering
\includegraphics[width=0.9\linewidth]{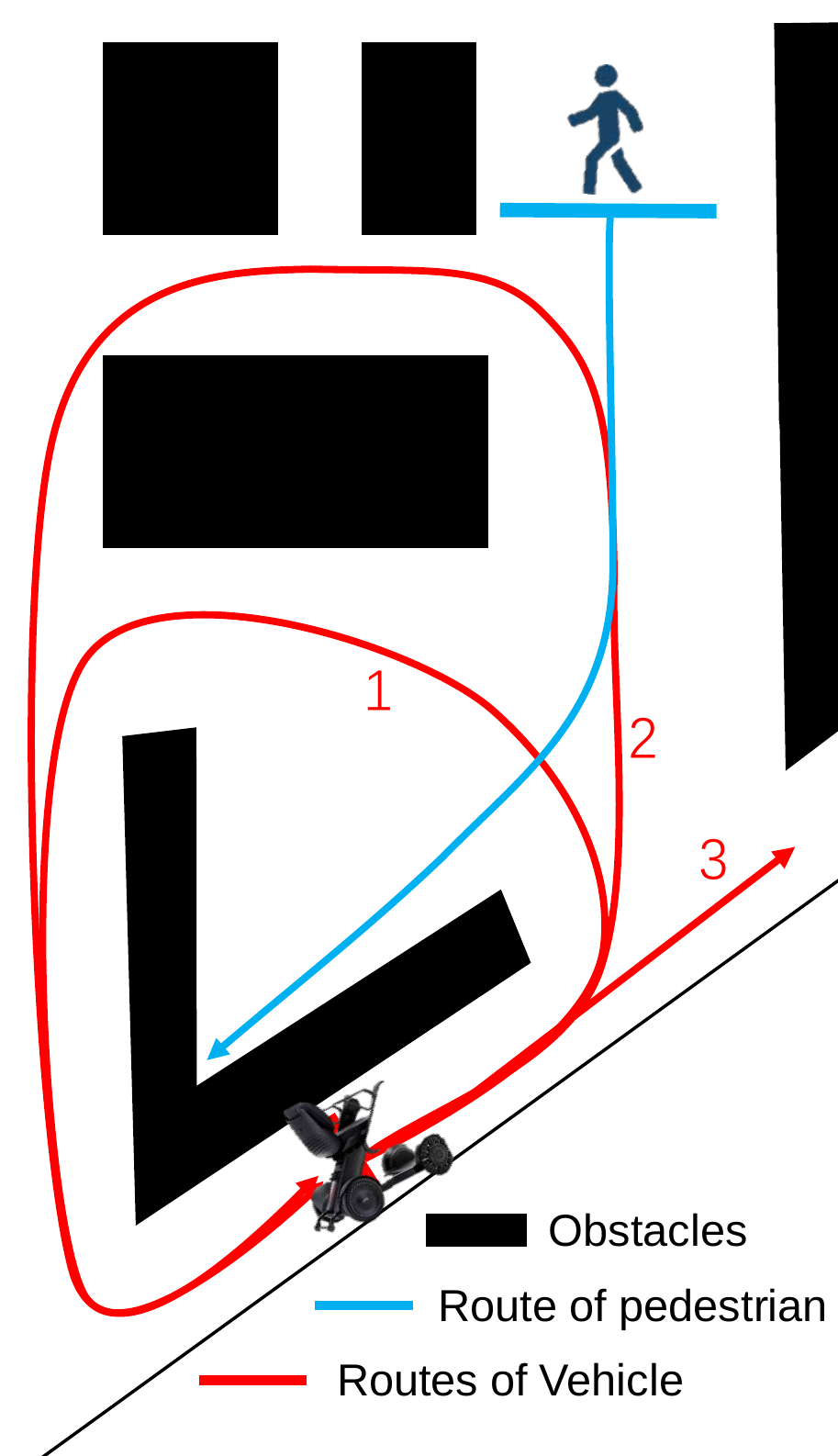}
\caption{Moving routes.}
\label{fig:map}
\end{minipage}
\begin{minipage}[t]{0.475\linewidth}
\centering
\includegraphics[width=0.9\linewidth]{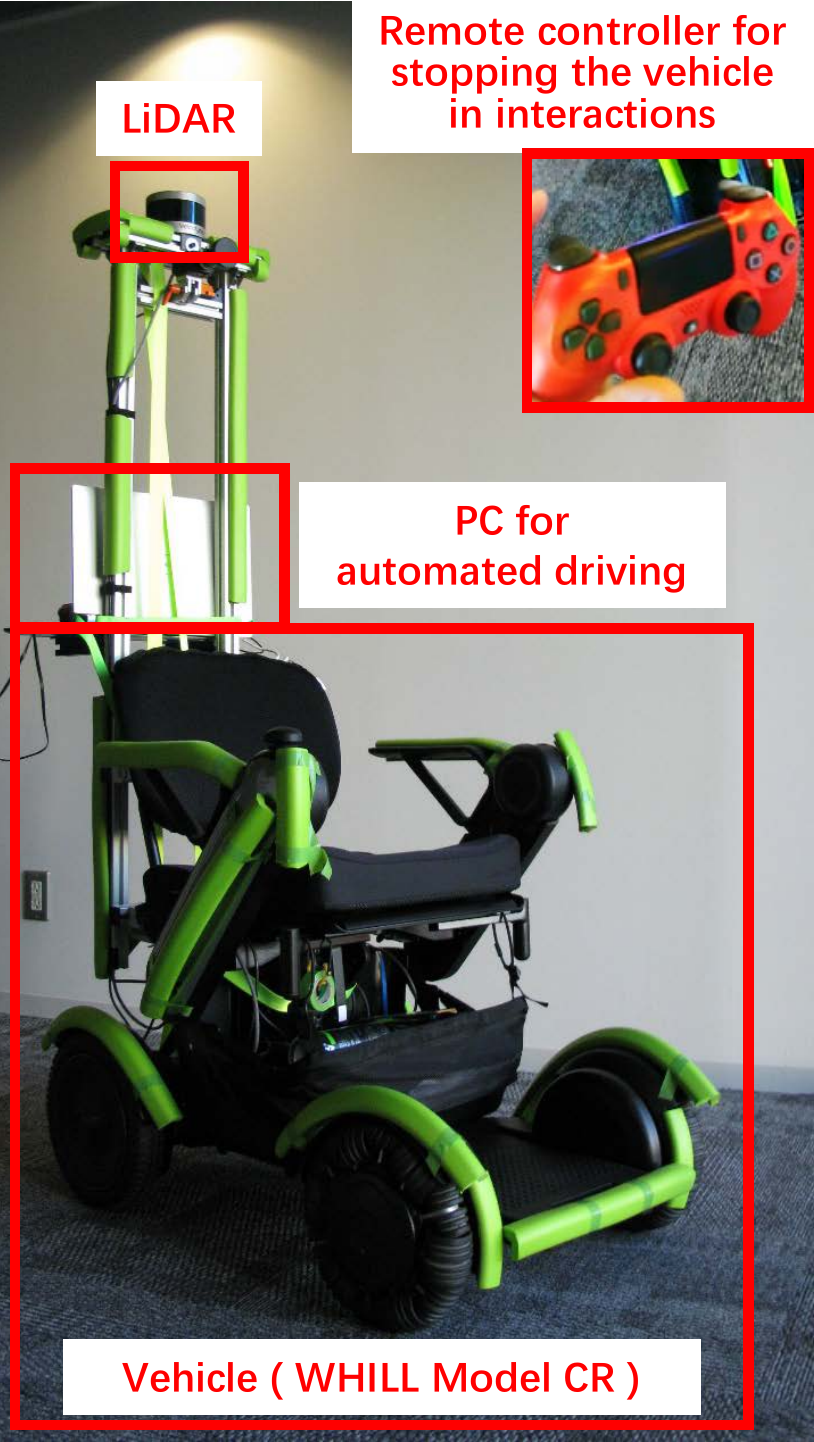}
\caption{Experimental vehicle.}
\label{fig:WHILL}
\end{minipage}
\vspace{-5mm}
\end{figure}

We convened 13 experimental participants within an age range of $20 \sim 29$ as pedestrians.
They had different educational backgrounds because they came from various disciplines of our university. 
All of them had no prior experience of interaction with AVs.
They were requested to walk from the origin to the destination as shown in Fig.~\ref{fig:map}.
They were told that a vehicle would interact with them during their walk.
Additionally, the participants were informed of interacting with a MV under the control of a driver and interacting with an AV automatically.
A wearable eye tracker {\it Tobii Pro Glasses 2} was used to measure the participants' gaze behavior during this experement.

A robotic wheelchair--{\it WHILL Model CR}~(Fig.~\ref{fig:WHILL}) was used as an experimental vehicle to interact with the participants during their walk.
The two modes (i.e., manual and automated) were used to drive the vehicle with a maximum speed of 1~[m/s]. 
In the manual driving mode, an experimenter rode on the vehicle and manipulated it using the available joystick.
The experimenter did not actively send information about their driving intention to the participants, but eye contact could not be ruled out.
In the automated driving mode, the vehicle was automatically controlled without a crew using a multi-layered LiDAR (Velodyne VLP--16) and wheel encoders. 
Importantly, the AV could not recognize participants, so it could not automatically interact with them.
To achieve a smooth interaction between the AV and the participants, a wireless remote controller was secretly used by the experimenter to control whether the AV would give the right-of-way to the participant.
In other words, when the AV automatically moved along the designed route, it stopped if the experimenter pressed a button on the remote control.
If the experimenter released the button on the remote control, the AV resumed the automatic movement.
The participants did not know that the AV was being manipulated by the experimenter.
For both the MV and AV, the experimenter adjusted the driving behavior through the distance to the participant, as well as the speed and walking direction of the participant, so as to realize interaction with the participant.

After the completion of each trial of interaction, the participants were required to complete a questionnaire on their subjective evaluations. 
There were two evaluation items on the questionnaire that were answered according to 5 point Likert scales.
The first question was used to evaluate the participants' understanding or confusion regarding the driving intention of the vehicle during the interaction according to the following scales:
{\it 1.~Completely did not understand}, 
{\it 2.~Did not understand much}, 
{\it 3.~Neutral},
{\it 4.~Mostly understood},
and {\it 5.~Fully understood}.
The second question was used to evaluate the participants' perception of safety during the interaction according to the following scales:
{\it 1.~Very dangerous},
{\it 2.~Slightly dangerous},
{\it 3.~Neutral},
{\it 4.~Slightly safe};
and {\it 5.~Very safe}.

There were three routes designed for the movement of the vehicle as shown in Fig.~\ref{fig:map}.
To simulate the scenario of a pedestrian interacting with the vehicle when crossing the street, route~1 was designed to cross the path of the participant.
In order to simulate the scenario of a pedestrian avoiding the vehicle, route~2 was designed to allow the vehicle and the participant move opposite each others on a straight road.
Route~3 was designed as a contrast, with no behavioral interaction between them.
Overall, each participant interacted with the MV and AV 20 times respectively.
Routes of those 40 interactions were chosen randomly.

This experiment was permitted by an ethics review committee of Institutes of Innovation for Future Society, Nagoya University.

\section{EXPERIMENT RESULTS}
\subsection{Data preprocessing}
Data of route 3 were excluded for analysis because this study focuses on the gaze behavior and psychological states of participants when interacting with the vehicle.
The observed gaze data of participant $\#11$ was also excluded because it had a lot of noise.
Besides, three trials of participant $\#3$'s gaze data for interacting with the MV were excluded due to equipment problems.
In total, data from 198 trials (route~1: 114 trials, route~2: 84 trials) of interacting with the MV and data from 204 trials (route~1: 120 trials, route~2: 84 trials) of interacting with the AV were observed.

{\it Tobii Pro Glasses 2} measured the foreground video and sequence of gaze points of each participant.
The size of the measured foreground video is $ 1920 \times 1080 $ pixels.
In each frame, the gaze point of the participant was recorded as a two-dimensional coordinate value on the plane of the foreground image.
In this experiment, the central visual field of participants was defined as a circle.
The gaze point was the center of a circle with a diameter of 108 pixels.
If any part of the vehicle area overlapped with part of the circle, then it was determined that the participant was gazing at the vehicle.
Under the abovementioned conditions, the total time of the gazes on the vehicle was calculated as the gaze duration in each trial.

The difference between this experiment and our previous experiment~\cite{liu2020gaze} is that each participant's gaze duration data was not standardized in this study.
The reason was that AVs cannot perceive the individual differences in the gaze duration of each participant in practical applications and situations.

\subsection{Subjective evaluation results}
Referring to the selection ratio of each subjective evaluation in Table~\ref{Tab:1}, the most frequent evaluation for the MV was {\it 5.~Fully understood} (42.4\%).
Besides, {\it 4.~Mostly understood} was the most frequent evaluation for the AV (44.1\%).
The selection ratio for the MV was significantly less than that for the AV when the participants chose {\it 1.~Completely did not understand} (MV is 1\%, AV is 3.4\%) and {\it 2.~Did not understand much} (MV is 4.5\%, AV is 19.1\%).
The above can be explained as being more difficult for them to understand the driving intention of the vehicle when interacting with the AV than when interacting with the MV.

Meanwhile, evaluation results for the perception of safety were similar to the evaluation results for the understanding of driving intentions.
Table~\ref{Tab:1} shows that the most frequently selected scale was {\it 5.~Very safe} for the MV (43.4\%) and {\it 4.~Slightly safe} for the AV (42.2\%).
0.5\% trails of interaction with the MV and 2.5\% trails of interaction with the AV were evaluated as {\it 1.~Very dangerous}.
The ratio of selecting {\it 2.~Slightly dangerous} when interacting with the AV (10.3\%) was also higher than that when interacting with the MV (8.1\%).
The above results show that participants felt situations to be more dangerous when interacting with the AV than when interacting with the MV.

\begin{table*}[bt]
\centering
\caption{Gaze durations and subjective evaluation scales.}
\vspace{-2mm}
\footnotesize
\setlength{\tabcolsep}{2mm}{
\begin{tabular}{l|rcc|rcc|c}
\hline
 & \multicolumn{3}{c|}{MV} & \multicolumn{3}{c|}{AV} & \multirow{2}{*}{\begin{tabular}[c]{@{}c@{}}Difference between median \\ values of gaze duration\\ on AV and MV {[}s{]}\end{tabular}} \\ \cline{2-7}
 & \begin{tabular}[c]{@{}c@{}}Selection\\ ratio\end{tabular} & \begin{tabular}[c]{@{}c@{}}Median\\ value {[}s{]}\end{tabular} & \begin{tabular}[c]{@{}c@{}}IQR\\ {[}s{]}\end{tabular} & \begin{tabular}[c]{@{}c@{}}Selection\\ ratio\end{tabular} & \begin{tabular}[c]{@{}c@{}}Median\\ value {[}s{]}\end{tabular} & \begin{tabular}[c]{@{}c@{}}IQR\\ {[}s{]}\end{tabular} &    \\ \hline
1. Completely did not understand & 1\%   & 3.249 & 0.929 & 3.4\%  & 6.156 & 2.329 & 2.907 \\
2. Did not understand much    & 4.5\%  & 3.058 & 2.459 & 19.1\%  & 4.478 & 4.058 & 1.420 \\
3. Neutral   & 15.2\%  & 2.750 & 1.901 & 21.1\% & 4.318 & 3.077 & 1.568 \\
4. Mostly understood  & 36.9\%  & 2.479 & 1.719 & 44.1\% & 3.885 & 2.867 & 1.406 \\
5. Fully understood   & 42.4\%  & 1.650 & 2.149 & 12.3\%  & 2.179 & 2.318 & 0.529 \\ \hline
1. Very dangerous    & 0.5\%  & 2.798 & 0.000 & 2.5\%  & 5.037 & 1.899 & 2.239 \\
2. Slightly dangerous  & 8.1\%  & 3.069 & 1.023 & 10.3\%  & 5.479 & 3.498 & 2.410 \\
3. Neutral   & 15.2\%  & 2.419 & 1.709 & 26.0\%   & 4.591 & 2.957 & 2.172 \\
4. Slightly safe & 32.8\%  & 2.799 & 2.119 & 42.2\%  & 3.703 & 2.773 & 0.904 \\
5. Very safe  & 43.4\%  & 1.535 & 2.019 & 19.1\%  & 2.219 & 2.379 & 0.684 \\ \hline
\end{tabular}
}
\label{Tab:1}
\vspace{-2mm}
\end{table*}

\subsection{Gaze durations for each subjective evaluation scale}

\begin{figure*}[t]
\centering
\begin{minipage}[t]{0.48\linewidth}
\centering
\includegraphics[width=1\linewidth]{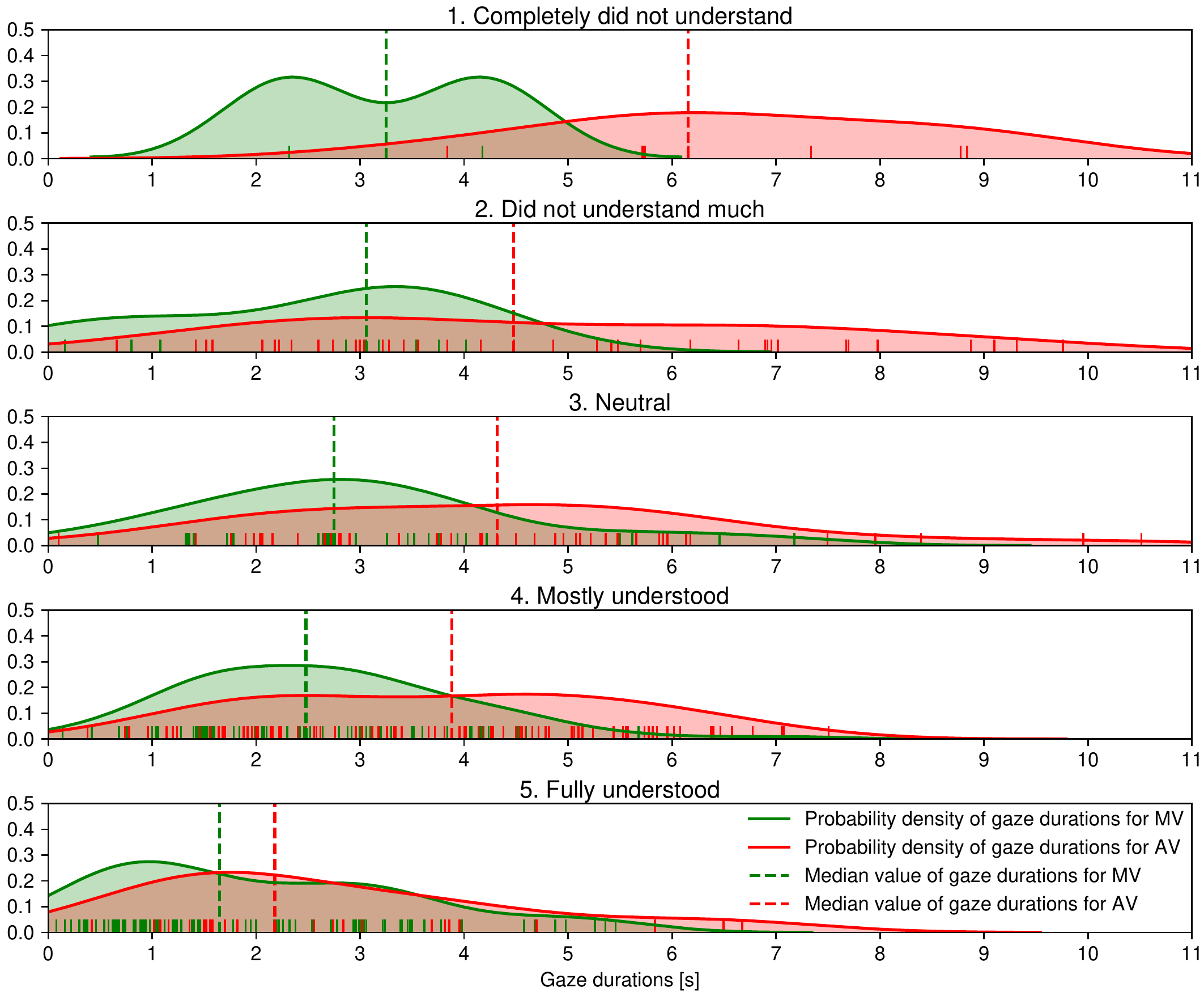}
\caption{Gaze durations for each subjective evaluation scale of understanding driving intentions.}
\label{fig:Gaze_under}
\end{minipage}
\hspace{2mm}
\centering
\begin{minipage}[t]{0.48\linewidth}
\centering
\includegraphics[width=1\linewidth]{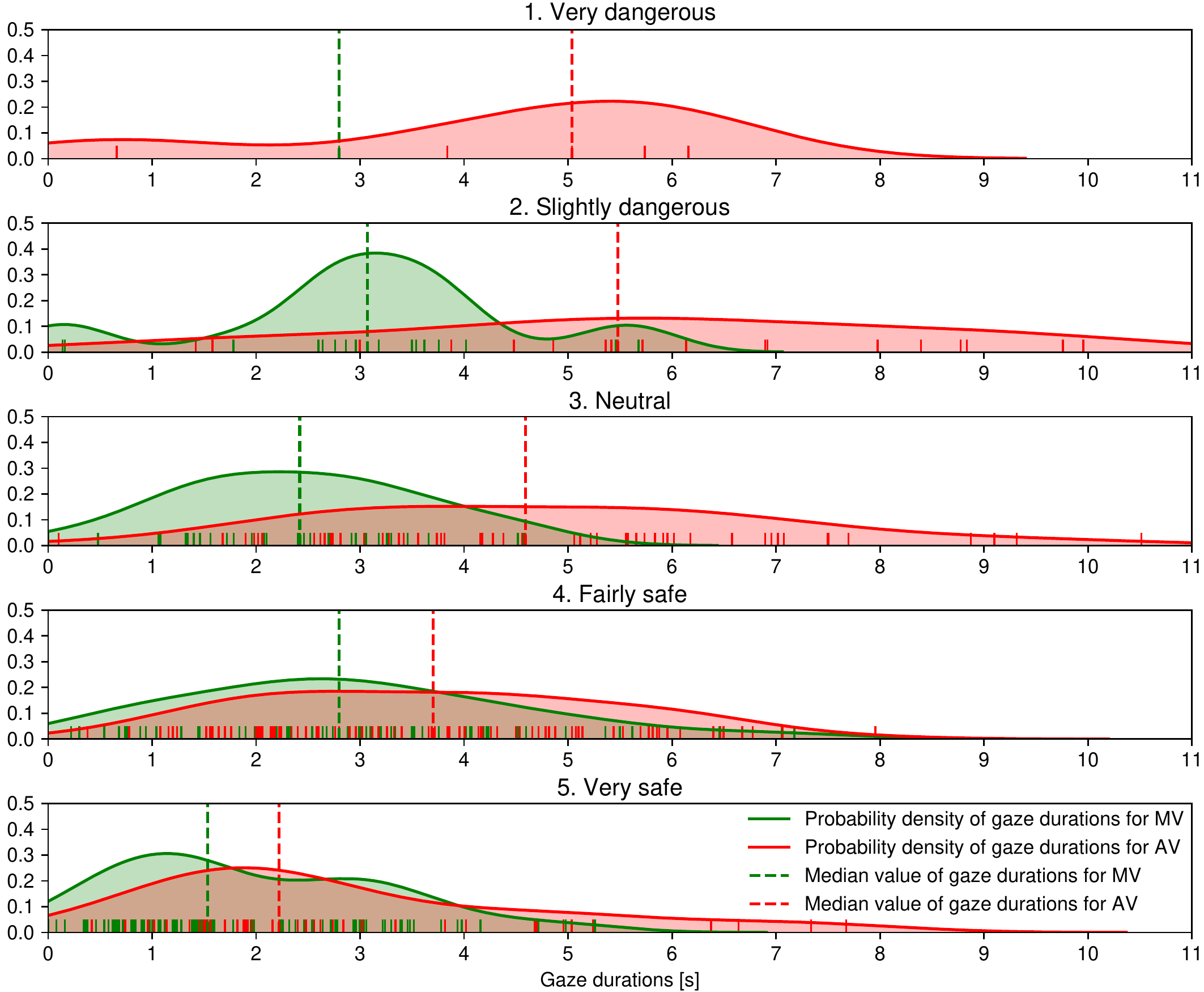}
\caption{Gaze durations for each subjective evaluation scale of the perception of safety.}
\label{fig:Gaze_safe}
\end{minipage}
\vspace{-6mm}
\end{figure*}

The differences in participant gaze durations on the MV and AV for different subjective evaluation scales were investigated.
Figs.~\ref{fig:Gaze_under} and \ref{fig:Gaze_safe} show the probability density of gaze durations for two subjective evaluations, i.e., for the understanding of driving intentions and for the perception of safety.
The vertical axis and the horizontal axis indicate the probability and the gaze duration, respectively.
Note that the horizontal axis signifies the integration time because the gaze duration is the total time of all gazes on the vehicle in the interaction.
In those graphs, the green color indicates interaction with the MV and the red color indicates interaction with the AV.
Samples of each gaze duration are represented by short vertical lines on the horizontal axis.
The median values are represented by long dotted lines.
To infer their probability density, kernel density estimation with a Gaussian kernel was used in order to account for individual differences that are potentially included in the gaze durations.
The inferred probability density of gaze durations for the MV and the AV are represented by green and red curves in Figs.~\ref{fig:Gaze_under} and \ref{fig:Gaze_safe}.

Accroding to Fig.~\ref{fig:Gaze_under} and Table~\ref{Tab:1}, the gaze durations for the AV were longer than that for the MV in terms of the median value of gaze durations corresponding to each scale of the evaluation for the understanding of driving intentions.
This showcased that if the participants did not understand the driving intention of the vehicle, then their gaze durations increased because they needed more time to observe the state of the vehicle and obtain information that could be used to infer the driving intention. 

Regarding each scale of the evaluation for the perception of safety, the median values of gaze durations on the AV were higher than on the MV, as shown in Fig.~\ref{fig:Gaze_safe} and Table~\ref{Tab:1}.
This shows that the gaze durations on the vehicle increased as the participants' perception of safety in interactions decreased.
It also implies that the participants watched the vehicle more attentively when they felt that it was dangerous.

Combining the results of these two subjective evaluations, we consider that the observation time (gaze duration) of the participants for the AV increases in order to prevent the AV becoming a danger to themselves when they did not understand the driving intention of the AV.

Meanwhile, for both the evaluations of the understanding of driving intentions and the perception of safety, the interquartile ranges~(IQR) of gaze durations on the AV were also higher than those on the MV, as shown in Table~\ref{Tab:1}.
This indicates that the participants had greater individual differences in their strategy of observation for the AV than for the MV because they did not have much experience interacting with the AV, especially in real world situations.

\begin{figure*}[t]
\centering
\begin{minipage}[t]{0.48\linewidth}
\centering
\includegraphics[width=1\linewidth]{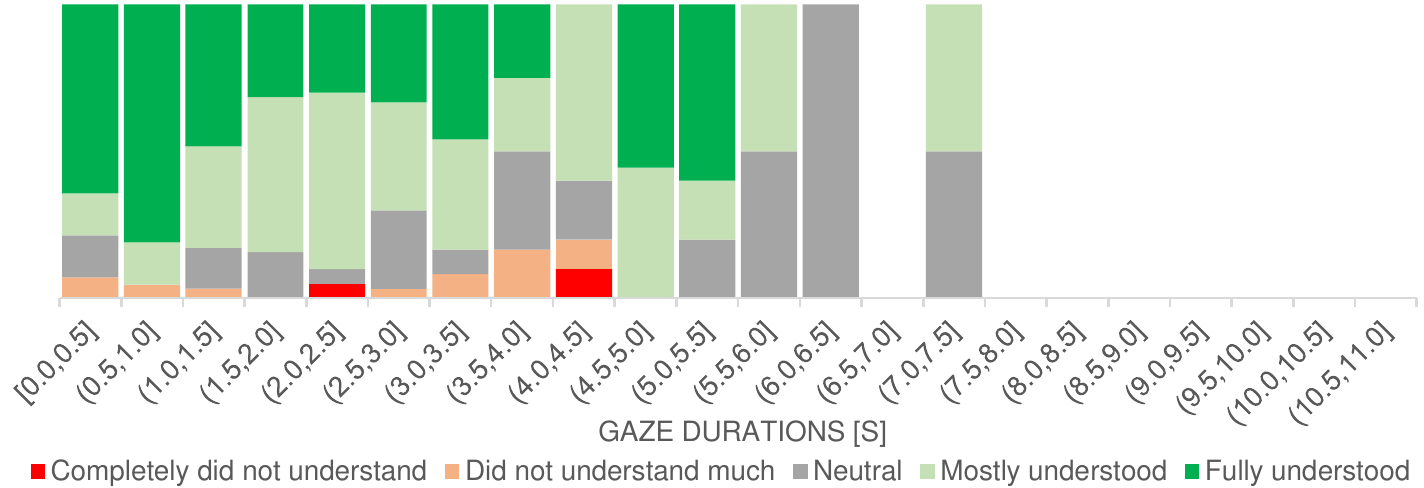}
\caption{Ratios of 5 point Likert scales for the understanding of driving intentions in every 0.5~[s] interval of gaze duration on the MV.}
\label{fig:Ratio_under_MV}
\vspace{2mm}
\centering
\includegraphics[width=1\linewidth]{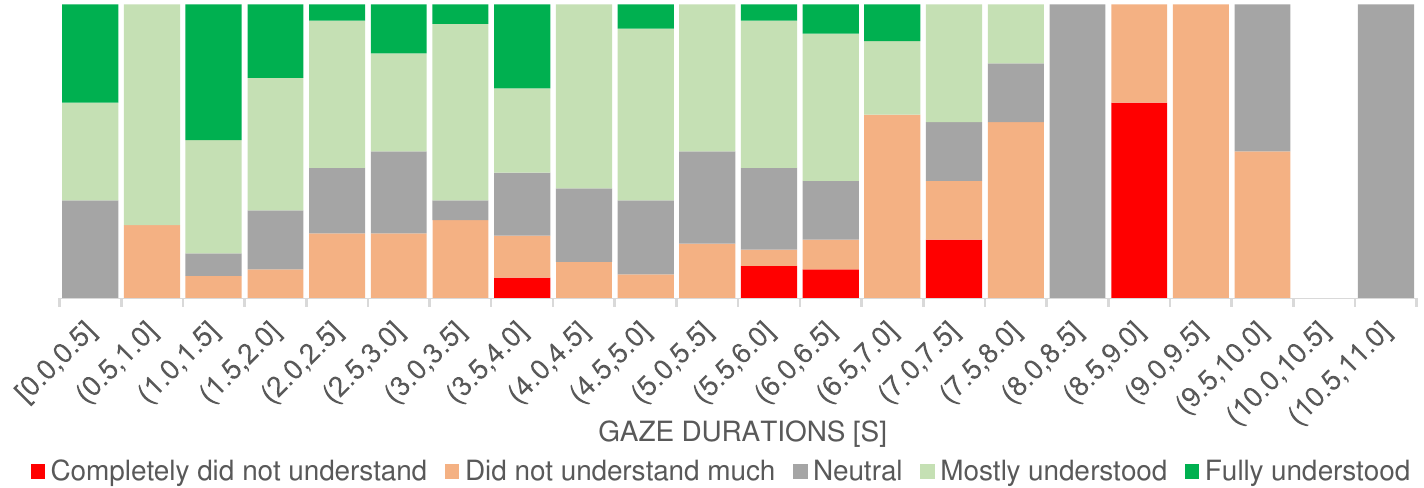}
\caption{Ratios of 5 point Likert scales for the understanding the driving intentions in every 0.5~[s] interval of gaze duration on the AV.}
\label{fig:Ratio_under_AV}
\end{minipage}
\hspace{2mm}
\centering
\begin{minipage}[t]{0.48\linewidth}
\centering
\includegraphics[width=1\linewidth]{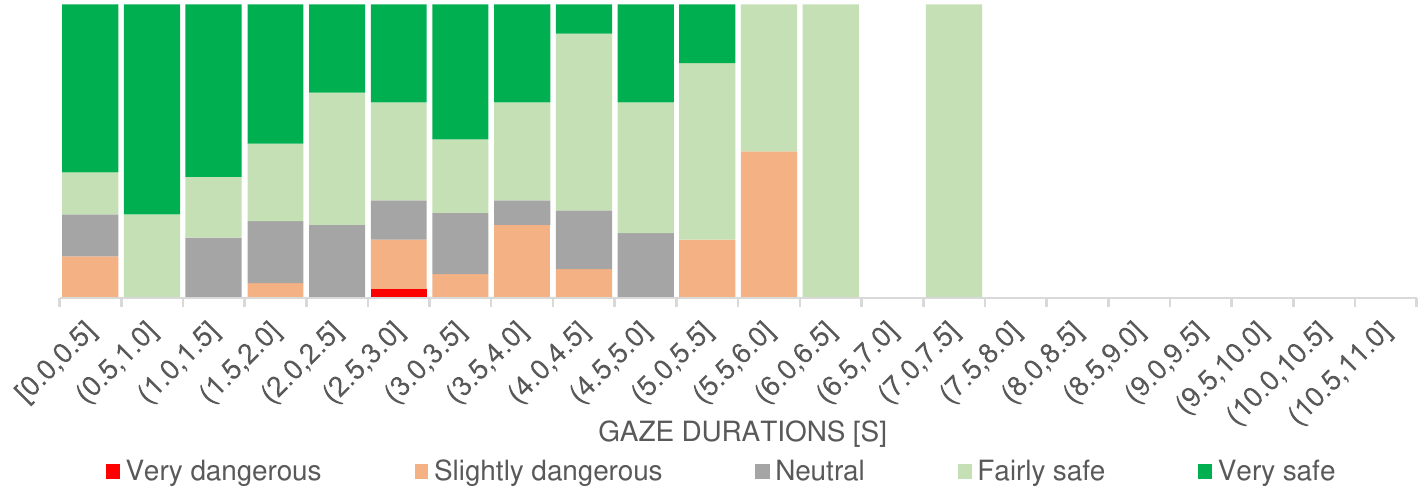}
\caption{Ratios of 5 point Likert scales for the perception of safety in every 0.5~[s] interval of gaze duration on the MV.\\}
\label{fig:Ratio_safe_MV}
\vspace{2mm}
\centering
\includegraphics[width=1\linewidth]{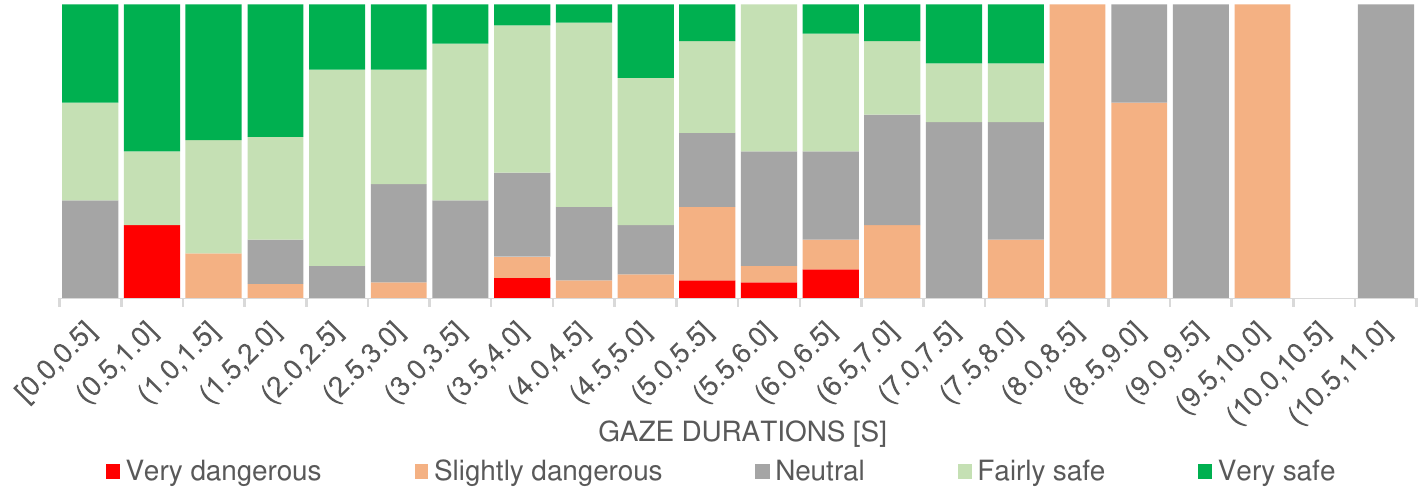}
\caption{Ratios of 5 point Likert scales for the perception of safety in every 0.5~[s] interval of gaze duration on the AV.}
\label{fig:Ratio_safe_AV}
\end{minipage}
\vspace{-7mm}
\end{figure*}

\subsection{What timing for an AV to make pedestrians understand its driving intentions after being noticed by them?}

The ratios of 5 point Likert scales for the understanding of driving intentions were checked for occurrance in every 0.5~[s] interval of gaze durations as shown in Figs.~\ref{fig:Ratio_under_MV} and \ref{fig:Ratio_under_AV}.
For the interaction between the participants and the MV, the ratios of {\it 4. Mostly understood} and {\it 5. Fully understood} were significantly higher than the others for most of the intervals, as shown in Fig.~\ref{fig:Ratio_under_MV}.
In contrast, for the interaction with the AV, the ratios of {\it 1. Completely did not understand} and {\it 2. Did not understand much}, increased, and the ratios of {\it 4. Mostly understood} and {\it 5. Fully understood} decreased as the gaze duration increased, as shown in Fig.~\ref{fig:Ratio_under_AV}.

\subsubsection{Lower bound of the timing} 
To determine the lower band of the timing that the AV should make the pedestrian understand its driving intentions after it is noticed,
we investigated the time range of gaze durations when the participants felt it difficult to understand the driving intention.
When the participants evaluated the AV's driving intention as {\it 2. Did not understand much}, the shortest gaze duration was more than $0.5$~[s].
Although it was a small ratio, it showed that the AV should make the pedestrian understand its driving intentions $0.5$~[s] after it is noticed, such as sending information about the driving intentions to the pedestrian.


\subsubsection{Upper bound of the timing} 
Each scale in the 5 point Likert scales has chance level of 20\% to be chosen.
Thus, there is a 40\% chance of choosing scales about ``did not understood'' (i.e., {\it 1. Completely did not understand} and {\it 2. Did not understand much}) or scales about ``understood'' (i.e., {\it 4. Mostly understood} and {\it 5. Fully understood}).
Therefore, a threshold for the evaluation ratio in 0.5~s intervals of gaze durations was set to 40\%.
Figure~\ref{fig:Ratio_under_AV} shows that only for interactions with the AV, the shortest gaze duration in the intervals that the participants did not understand the driving intentions in more than 40\% trials was over 6.5~[s]. 
Therefore, we recommend that the AV is better to make the pedestrian understand its driving intentions accurately within the first 6.5~[s] while the pedestrian is gazing at it.

\subsection{What timing for an AV to make pedestrians feel safe after being noticed by them?}
We calculated the ratios of 5 point Likert scales for the perception of safety for occurrance in every 0.5~[s] interval of gaze durations as shown in Figs.~\ref{fig:Ratio_safe_MV} and \ref{fig:Ratio_safe_AV}.
As shown in Fig.~\ref{fig:Ratio_safe_MV}, for the evaluation result of interacting with the MV, {\it 4.~Slightly safe} and {\it 5.~Very safe} were chosen in most of the intervals.
For the interaction with the AV, the ratios of {\it 1.~Very dangerous} and {\it 2.~Slightly dangerous} increased, and the ratios of {\it 4. Slightly safe} and {\it 5.~Very safe} decreased as the gaze duration increased, as shown in Fig.~\ref{fig:Ratio_safe_AV}.


\subsubsection{Lower bound of the timing} 
We also investigated the shortest gaze duration when the pedestrians felt dangerous to determine the lower band of the timing that the AV should make the pedestrian feel safe after it is noticed.
Fig.~\ref{fig:Ratio_safe_AV} shows that the shortest gaze duration when the participants chose {\it 1.~Very dangerous} was in the interval $0.5 \sim 1.0$~[s].
Thus, the AV is better to start to make the pedestrian feel safe 0.5~[s] after it is noticed.

\subsubsection{Upper bound of the timing} 
Similarly, the total of {\it 4. Slightly safe} and {\it 5. Very safe} has a chance level of 40\% to be chosen.
Thus, a threshold for the evaluation ratio of the perception of safety was set to 40\%.
Referring to Fig.~\ref{fig:Ratio_safe_AV}, there were high ratios where the participants felt danger when interacting with the AV for their gaze durations of over 8.0~[s].
In particular, if the gaze duration was more than 8 seconds, there were no cases that were evaluated as {\it 4. Slightly safe} or {\it 5. Very safe}.
In other words, the AV is more likely to enable the pedestrian to feel safe within the first 8.0~[s] while the pedestrian is gazing at it, e.g., sending related information to the pedestrian.


\subsection{Discussion}
We found a trend through Fig.~\ref{fig:Gaze_under} that the increase in gaze durations on the AV was gradually greater than the gaze durations on the MV as the participants understood the driving intention of the vehicle less.
For example, there was no clear difference between the probability densities of the gaze durations on the AV and the MV when the participants chose {\it 5.~Fully understood}, but the difference for {\it 1.~Completely did not understand} was large, as shown in Fig.~\ref{fig:Gaze_under}.
Referring to the results of the perception of safety in Fig.~\ref{fig:Gaze_safe}, it also had the same trend.
Those trends are also shown in Table~\ref{Tab:1} as the differences between median values of gaze durations on the AV and the MV for each evaluation scale.
The reason for this trend is speculated that the participants sufficiently trust in both AV and MV when they understood the vehicle's driving intentions and felt safe.
As their understanding of driving intentions becomes ambiguous and the perception of danger increases, their trust in the vehicle may decline, while they increase the observation time of the vehicle to gain more information to protect themselves from danger.
An important factor to be considerd here is the participants potentially trust in the driver.
We speculate that this is also one of the reasons that the participants' gaze duration on the MV was shorter than that on the AV when they understood the vehicle's driving intentions and felt safe.

Besides, in Figs.~\ref{fig:Ratio_under_MV} and \ref{fig:Ratio_under_AV}, we found that when the participants interacted with the MV, the earliest occurrences of {\it 4. Mostly understood} and {\it 5. Fully understood} were earlier than when they interacted with the AV.
When the participants only observed the MV $0.0 \sim 0.5$~[s] during the interaction, there were cases that the MV's driving intention was evaluated as {\it 2. Did not understand much}.
The same difference occurred in {\it 1. Completely did not understand}, where the earliest occurrences were at $2.0 \sim 2.5$~[s] for the MV and $3.5 \sim 4.0$~[s] for the AV.
Based on the above results, we consider that the participants trusted the MV more than the AV because they successfully performed the trials even with a brief observation of the MV or unclear understanding of driving intentions of the MV.

On the basis of the above discussions, we assume that the underlying cause of this may be related to the pedestrians having less trust in the AV than in the human driver of the MV.
This may be the key factor that makes it difficult for AVs to achieve popularity in society.

\section{CONCLUSION}
We designed an experiment of pedestrian--vehicle interaction to find a time range that an AV should make a pedestrian understand its driving intentions and feel safe in an interaction.
Thirteen participants were invited to interact with the MV and the AV.
The participants' gaze information and their subjective evaluation for the understanding of driving intentions as well as their perception of safety were collected.
By analyzing the participants' gaze duration on the vehicle with their subjective evaluations, we found that
1) the AV is better to enable the pedestrian to understand its driving intentions accurately within $0.5 \sim 6.5$~[s] while the pedestrian is gazing at it, and
2) the AV is better to enable the pedestrian to feel safe within $0.5 \sim 8.0$~[s] while the pedestrian is gazing at it.

A limitation of this paper is that the experimental results of this study can be applied to personal mobility vehicles or mobile robots running at low speed, but it may be difficult to apply them to an automated car running at high speed.
The reason is that the gaze duration of participants may depend on the speed of the vehicle.

In future, we will use the same experimental method to study the interaction between pedestrians and an automated car.
We will also focus on establishing an eHMI for AVs that can quickly, clearly, and kindly convey driving intention to pedestrians, thus improving the acceptability of AVs.
Additionally, the results of this study indicate that the gaze behavior of pedestrians on the AV may depends on the trust that the pedestrians have.
Besides, some related work has also focused on the feeling of user safety and trust in using AV systems~\cite{mcallister2017concrete,liu2019driving,hollander2019overtrust}. 
User trust has to be taken into account in our future study.

\vspace{-2mm}
\section*{ACKNOWLEDGMENT}
\vspace{-2mm}
This work was supported by JST-Mirai Program Grant Number JPMJMI17C6, and JSPS KAKENHI Grant Numbers JP19K12080, Japan.
\vspace{-2mm}

\bibliographystyle{ieeetr}
\bibliography{sample-base}

\end{document}